\documentclass[10pt, twocolumn]{IEEEtran}

\usepackage{xspace,amsmath,amssymb,epsfig,syntonly,cite}
\usepackage{verbatim}

\newtheorem{algorithm}{\hspace{-11pt}\bf Algorithm}

\usepackage{epstopdf}
\usepackage{color}
\usepackage{amsmath,amssymb}
\usepackage{lettrine}
\usepackage{autobreak}

\newcommand*{\blue}{\color{black}}

\makeatletter
\let\myorg@bibitem\bibitem
\def\bibitem#1#2\par{%
 \@ifundefined{bibitem@#1}{%
   \myorg@bibitem{#1}#2\par
 }{%
   \begingroup
     \color{\csname bibitem@#1\endcsname}%
     \myorg@bibitem{#1}#2\par
   \endgroup
 }%
}

\makeatother

\makeatletter
\newif\if@restonecol
\makeatother

\usepackage[linesnumbered,ruled,lined,boxed]{algorithm2e}
\usepackage{algpseudocode}
\usepackage{amsmath}
\SetKwRepeat{Do}{do}{while}
\long\def\symbolfootnote[#1]#2{\begingroup
\def\thefootnote{\fnsymbol{footnote}}
\footnote[#1]{#2}\endgroup}

\psfull

\begin{document}
\title{Detection and Mitigation of Position Spoofing Attacks on Cooperative UAV Swarm Formations}
\author{
   {Siguo Bi},
   {Kai~Li},~\IEEEmembership{Senior Member,~IEEE},
   {Shuyan~Hu},~\IEEEmembership{Member,~IEEE},\\
    {Wei~Ni},~\IEEEmembership{Fellow,~IEEE},
   {Cong~Wang}, 
    and {Xin~Wang},~\IEEEmembership{Fellow,~IEEE}

    \thanks{
Manuscript received July 17, 2023; revised October 18, 2023; accepted December 3, 2023. Work in this paper was supported by the National Natural Science Foundation of China under Grants No. 62231010, No. 62071126, and No. 62101135, and the Innovation Program of Shanghai Municipal Science and Technology
Commission Grant No. 21XD1400300. The work of K. Li was supported by the CISTER Research Unit (UIDP/UIDB/04234/2020) and project ADANET (PTDC/EEICOM/3362/2021), financed by National Funds through FCT/MCTES (Portuguese Foundation for Science and Technology).

S. Bi, S. Hu, C. Wang, and X. Wang are with Fudan University, Shanghai, China (emails:~\{fdbsg, syhu14, congwang, xwang11\}@fudan.edu.cn).

K. Li is with the Division of Electrical Engineering, Department of Engineering, University of Cambridge, CB3 0FA Cambridge, U.K., and also with Real-Time and Embedded Computing Systems Research Centre (CISTER), Porto 4249--015, Portugal (E-mail: kaili@ieee.org).

W. Ni is with the Commonwealth Scientific and Industrial Research Organization (CSIRO), Sydney, NSW 2122, Australia (e-mail: wei.ni@data61.csiro.au).

Corresponding author: X. Wang.
}}


\maketitle


\begin{abstract}
Detecting spoofing attacks on the positions of unmanned aerial vehicles (UAVs) within a swarm is challenging. Traditional methods relying solely on individually reported positions and pairwise distance measurements are ineffective in identifying the misbehavior of malicious UAVs. This paper presents a novel systematic structure designed to detect and mitigate spoofing attacks in UAV swarms.
We formulate the problem of detecting malicious UAVs as a localization feasibility problem, leveraging the reported positions and distance measurements. To address this problem, we develop a semidefinite relaxation (SDR) approach, which reformulates the non-convex localization problem into a convex and tractable semidefinite program (SDP). Additionally, we propose two innovative algorithms that leverage the proximity of neighboring UAVs to identify malicious UAVs effectively.
Simulations demonstrate the superior performance of our proposed approaches compared to existing benchmarks. Our methods exhibit robustness across various swarm networks, showcasing their effectiveness in detecting and mitigating spoofing attacks. {\blue Specifically, the detection success rate is improved by up to 65\%, 55\%, and 51\% against distributed, collusion, and mixed attacks, respectively, compared to the benchmarks.}
\end{abstract}

\maketitle

\setcounter{page}{1}

\begin{keywords}
 Malicious UAV detection, position spoofing attack, cooperative localization, semidefinite programming.
\end{keywords}

\section{Introduction}
\begin{figure}[t]
 \center{\includegraphics[width=7.5cm]  {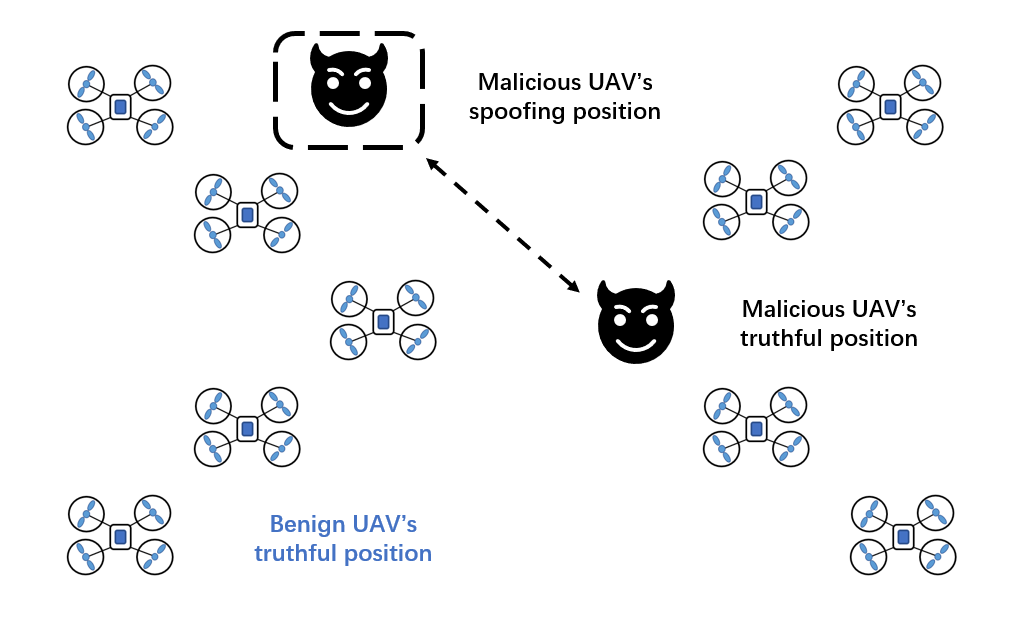}}
 \caption{\label{1} An illustration of the attack model, where a malicious UAV falsifies its position and broadcasts the fake position to the other benign UAVs.}
 \label{F:schema}
 \end{figure}

Recently, there has been a widespread utilization of unmanned aerial vehicles (UAVs)~\cite{hu21}, including parcel delivery~\cite{DBLP:journals/tits/LiuNLGZ23a}, {\blue radio surveillance~\cite{yuantifs,savkin2023effective}}, and rescue missions~\cite{kli19}. This is due to the affordability and endurance of UAVs, and {\blue their flexibly adjustable positions conducive to line-of-sight (LOS) communications}, facilitated by rapid technological advancements~\cite{hu22}.
{\blue The varieties of practical needs for UAV swarms further ignite and necessitate the protection of security for UAV swarms~\cite{tifsy}.
For instance, the reliability of information propagation has been analyzed in large-scale networks, including UAV swarms~\cite{Songtifs}. 
The connectivity of a UAV swarm has been studied in the presence of jamming attacks from the ground~\cite{DBLP:journals/tvt/YuanFNWLX20}.}
To enhance reliability and mitigate potential flight collisions, it is crucial to establish a formation flight and coordination among UAVs~\cite{Bi2023}. 
In the formation flight of a UAV swarm, individual UAVs rely on position reports from their peers and their pairwise distance measurements with neighboring UAVs to maintain inter-UAV distances and avoid collisions. Compromised or malicious UAVs can launch position spoofing attacks, potentially leading to catastrophic consequences for the UAV swarm \cite{Yao22}. A malicious UAV might transmit a deceptive position report, misleading other UAVs while simultaneously concealing its true location, as illustrated in Fig. \ref{F:schema}. 
Such conditions can disrupt the control mechanism that maintains swarm formation, resulting in disorders \cite{Choi18}. 

Detecting and identifying malicious UAVs within a swarm is challenging. 
This problem seeks to establish whether a feasible position realization for each UAV that aligns with all reported distances and measurements exists. 
Such a feasibility problem is non-trivial and non-convex, and has never been studied and addressed in the existing literature. 
As delineated in this paper, the problem can be transformed into a convex semidefinite program (SDP), allowing for efficient use of
convex optimization solvers in polynomial time.
However, solving the SDP problem alone does not enable the identification of individual malicious UAVs. 
On the one hand,  the number of malicious UAVs is typically unknown and needs to be detected. 
On the other hand, the effectiveness of SDP can be penalized by the interdependence among the positions of neighboring UAVs. 

To address these challenges and precisely identify malicious UAVs, this paper proposes two new algorithms: the Cooperative Detection and Identification (CDI) algorithm and the Enhanced CDI (E-CDI) algorithm. 
The CDI algorithm initiates its process by creating sets of possible malicious and benign UAVs. Subsequently, it combines the selected potentially malicious UAVs with the benign set, establishing a connected sub-network for the SDP-based position feasibility check. If all the neighboring UAVs of a selected UAV are themselves malicious, the CDI algorithm may misjudge the UAV as malicious, as attempting to localize a sub-network with an entire malicious neighborhood is inherently unfeasible.
{\blue In contrast}, the E-CDI algorithm conducts an additional localization feasibility check on each individual UAV in the neighborhood, compared to the CDI algorithm. 
By this means, collusion attacks launched by multiple closely located, malicious UAVs can be detected and mitigated.


{\blue Compared to the existing relevant works, e.g., \cite{Yao22,Buske2022,Liu202011756,Choi18}, the new contributions of this paper include:
\begin{enumerate}
\item To detect position spoofing attacks, we propose a novel mechanism for malicious UAV detection and identification, where we cast the challenging malicious UAV detection problem as a localization feasibility problem. 
\item A semidefinite relaxation (SDR) approach is put forth to transform the non-convex feasibility problem into a convex problem. The presence of malicious UAVs can then be efficiently ascertained by evaluating the feasibility of the convex problem.
\item We develop two iterative algorithms, i.e., CDI and E-CDI, to identify malicious UAVs by leveraging the proximity of neighboring UAVs. 
\begin{itemize}
    \item 
The CDI algorithm dynamically merges selected potentially malicious UAVs into the benign set to form a connected positioning sub-network. This sub-network is used to determine whether the selected UAV is malicious.
    \item 
The E-CDI algorithm enhances identification efficiency by further assessing each neighboring UAV in the neighborhood of a potentially malicious UAV. As a result, collusion attacks launched by multiple closely located, malicious UAVs can be detected.   
\end{itemize}
Both algorithms are designed to conclude within a finite number of iterations and exhibit robust performance across various network configurations of UAV swarms.
\end{enumerate}
Extensive simulations demonstrate that the proposed CDI and E-CDI algorithms achieve superior performance on classic metrics compared to the benchmark techniques.}
{\blue Under the proposed algorithms, the detection success rate can be improved by up to 65\%, 55\%, and 51\% against distributed, collusion, and mixed attacks, respectively, compared to their benchmarks.}

The rest of this paper is organized as follows. Section II reviews the related works. Section III formulates and convexifies the malicious UAV’s misbehavior detection problem. In Section IV, two efficient iterative algorithms are proposed to identify malicious UAVs. Section V provides numerical results to evaluate the proposed algorithm, followed by conclusions in Section VI. 

{\color{black}
\textit{Notation:} 
Upper- and lower-case boldface symbols denote matrices and vectors, respectively; 
$|\cdot|$ takes the absolute value if a scalar is concerned or the cardinality if a set is concerned; $\left\|\cdot \right\| $ denotes $\ell_2$-norm; $\hat{(\cdot)}$ indicates a reported, noise-corrupted version of $(\cdot)$. 
The notation used is collated in Tab.~\ref{table_symbols}.
}

\begin{table}
\centering 
\caption{\color{black} Notation and definition.}
\label{table_symbols}
\begin{tabular}{p{0.75cm}  p{6.9cm}} 
\hline
 \textbf{Notation} & \textbf{Definition}   \\ \hline 

  ${\cal X}$ & The set of the 3D coordinates
of all the UAVs \\ 
$N$ & The total number of UAVs   \\ 

$\boldsymbol{x}_{i}$ & The actual position of the $i$-th UAV   \\ 

$\hat{\boldsymbol{x}}_{i}$ & The reported position of the $i$-th UAV   \\ 

 $r_{ij}$ & The actual distance  between UAVs~$i$ and $j$, $i\neq j$   \\ 

 $\hat{r}_{ij}$ & The reported distance between UAVs~$i$ and $j$, $i\neq j$   \\ 

 $\hat{\alpha}_{ij}$ & An auxiliary
variable
\\ 

 $\boldsymbol{w}_{i}$ & The noise vector for position measurement of UAV $i$ \\ 

 $w_{ij}$ & The noise in the reported distance measurement between UAVs~$i$ and~$j$, $i\neq j$ \\ 
 $\boldsymbol{I}_{3}$  & The $3\times 3$ identity matrix \\ 

 $\boldsymbol{X}$  & The $3 \times N$ matrix  with its $i$-th column being $\boldsymbol{x}_{i}$ \\ 


   $d$ & The communication range for distance measurement   \\ 
   $\epsilon$ & A small constant, e.g., $1\times10^{-6}$  \\ 
  $\boldsymbol{e}_{i}$ & The vector whose $i$-th element is one and
the rest are zeros.  \\ 

 $\rho_{ij}$ & The indicator of whether UAVs $i$ and $j$ are directly connected.   \\ 

      $\boldsymbol{E}_{n}$ & The matrix of the measured and reported Euclidean distances between directly connected UAVs
      \\ 

            $\boldsymbol{E}_{r}$ & The matrix of the Euclidean distances between directly connected UAVs generated based on the reported positions of the UAVs  \\ 
 $\mathbb{N}$ & The set of all $N$ UAVs.   \\ 
 
   $\mathbb{M}$ & The set of malicious
UAVs  \\ 

  $\mathbb{B}$ & The set of benign UAVs   \\ 

   $\mathbb{N}_{k}$ & The set of the one-hop neighbors of UAV $k$  \\ 



            $R_M$  &  The malicious ratio, i.e., the ratio of the number of malicious UAVs to the total number of UAVs in a UAV swarm      \\ \hline

\end{tabular}
\end{table}

\section{Related Work}
\subsection{Spoofing Attacks on UAVs}
Spoofing attacks on UAVs have been extensively investigated in the recent literature.
However, most works have focused on the direct hijack of the Global Positioning System (GPS) of a specific single UAV. Aiming at identifying fake GPS coordinates due to the hijack of the GPS communication software, the authors of \cite{Jiang2022791} proposed a convolutional neural network (CNN) integrated with a recurrent neural network (RNN) to predict a vehicle's real-time trajectory based on the data from multiple sensors.
With a similar purpose of handling GPS spoofing attacks, the authors of \cite{Pardhasaradhi202211122} proposed a two-step approach based on data sensed and fused from distributed radar ground stations equipped with a local tracker. The approach consists of spoofing detection and mitigation. In the spoofing detection step, a track-to-track association approach was adopted to detect spoofing attacks with fused data from UAVs and a local tracker. In the mitigation step, the fused data was input to a controller to mitigate the spoofing attack detected. The proposed two-step approach was reported to achieve almost the same accuracy as GPS efficiently.

To enhance the reliability of flight controllers when the UAV is under GPS spoofing attack, the authors of \cite{10027832} utilized an extended Kalman filter (EKF)-based approach.
They investigated the impact of GPS spoofing on the EKF estimation and the UAV itself under different levels of attack strength. It was reported that the classic EKF-based approach can tolerate small errors from spoofing attacks, but can be inefficient when the attack intensifies.
Similar works on GPS-related spoofing attacks on a specific single UAV can be found in \cite{9419417,9824124,9700664,Buske2022}, and spoofing attacks related to the time-of-arrival (TOA) or time difference-of-arrival (TDOA) can be found in \cite{Wang20221ssss}.

The security issue of UAV swarms has attracted increasing attention. In order to mitigate the navigation spoofing attacks on aerial formations, the authors of \cite{Michieletto20221} proposed a cascaded estimation algorithm used for concurrent GPS spoofing detection and localization. An attack detection module was based on the consistency of multimodal measurement to realize threshold tests. A localization module was then used for a decision based on remarkable differences between safe and under-attack conditions of UAV self-localization. The cascaded approach can achieve a safe self-localization for a UAV swarm under a spoofing attack. Aiming at solving the GPS spoofing attack in a UAV swarm, the authors of \cite{Babaghayou2022343} proposed a  security-aware monitoring method to monitor the potential malicious UAVs and protect the benign ones from attacks. The method was implemented by the received-signal-strength-indicator (RSSI)-based triangulation. 

\subsection{Cooperative Network Localization}
Position-related spoofing attacks destroy the localization of UAVs in a UAV swarm, since a UAV swarm can be considered a cooperative network. 
SDP, an efficient convex optimization approach, has been extensively applied to cooperative network localization. Employing the SDP, the authors of \cite{bi_iot2023} proposed a novel difference-of-convex (DC)-based algorithm to achieve accurate cooperative localization.
The authors of \cite{Wang20232142} proposed an SDP-based method to estimate the relative transformation of a robot in a cooperative robotic swarm. The SDP-based method could achieve global optimality and scalability. The authors of \cite{Yang2022} developed an efficient SDP-based scheduling strategy to optimize UAV deployment in intelligent transportation cooperative networks. More SDP-based cooperative localization techniques can be found in~\cite{Liu202011756,Jiang20201903,Wu201998}.

Unlike existing works that rely heavily on extensive training using historical data, we put forth an SDP-based UAV misbehavior detection mechanism with no need for historical data. The proposed SDP-based mechanism detects and identifies malicious UAVs that misreport their positions by leveraging the proximity of neighboring UAVs. This mechanism is applicable regardless of the specific type of localization signals hijacked and spoofed, including GPS, TDA, or TDOA.

\section{System Model and Problem Formulation}

In this section, we first provide the threat model considered and formally formulate the malicious UAV detection problem as a localization feasibility problem. By applying the SDR, we convexify the feasibility problem into a convex problem. 

\subsection{Threat Model}
  \begin{figure} 
 \center{\includegraphics[width=7.5cm]  {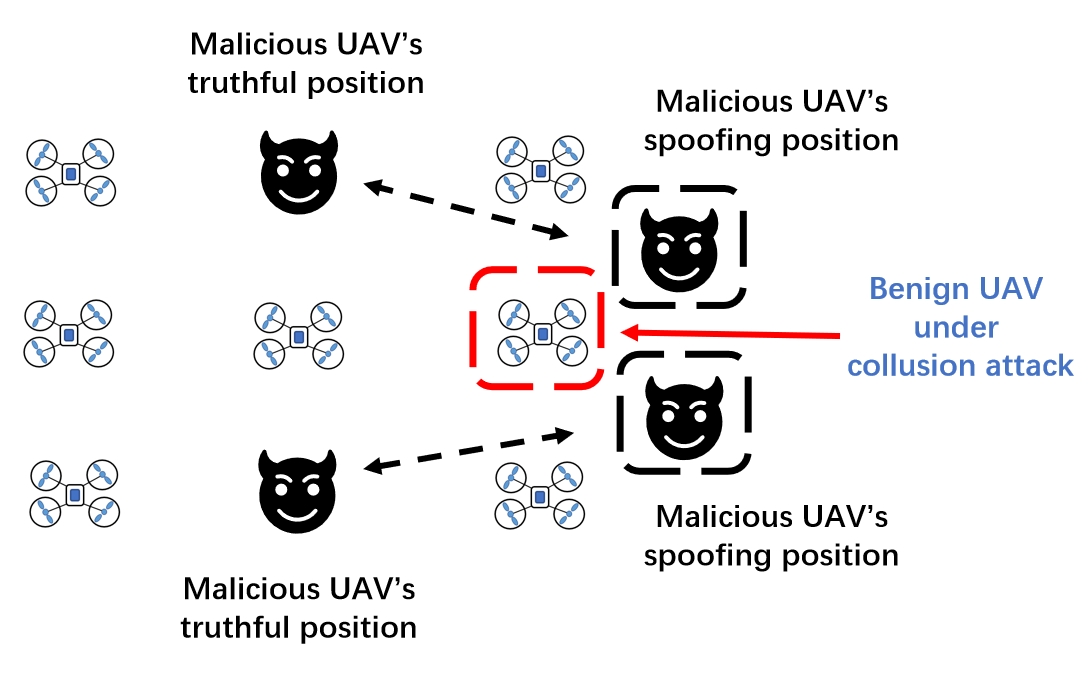}}
 \caption{\label{1} {\blue An illustration of the collusion spoofing attack model, where two malicious UAVs falsify their positions to be within the one-hop neighborhood of the benign UAV.} }
 \label{F:model_collusive}
 \end{figure}

  \begin{figure} 
 \center{\includegraphics[width=7.5cm]  {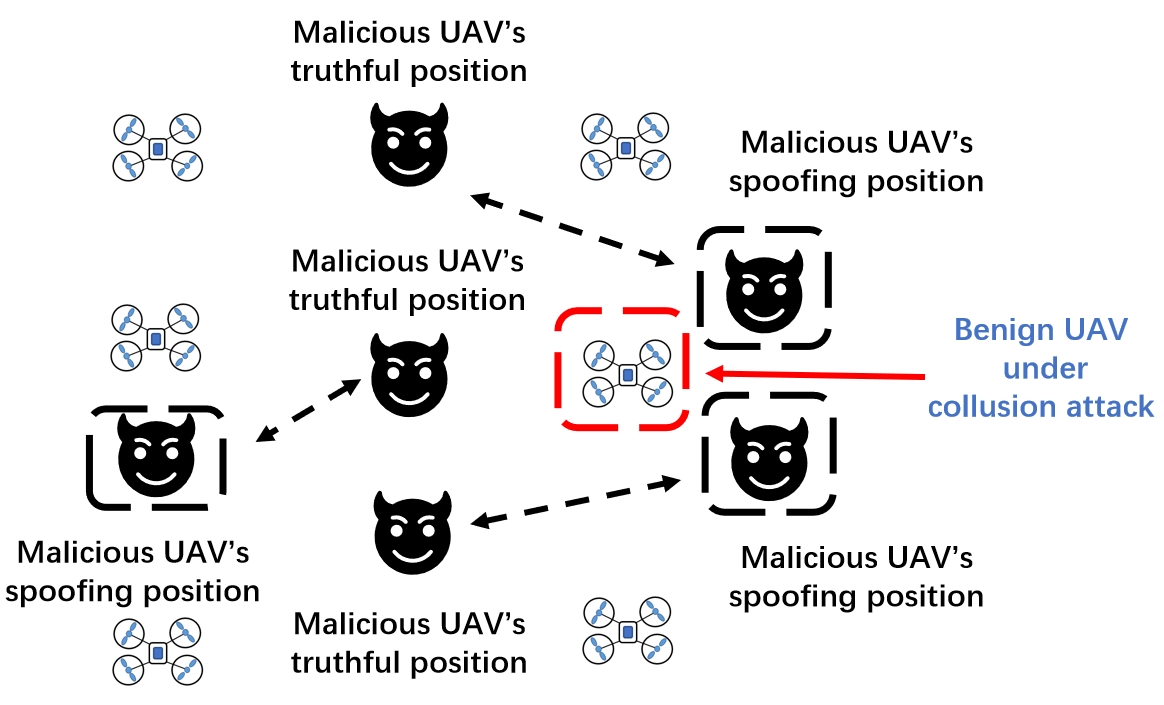}}
 \caption{\label{1} {\blue An illustration of the mixed spoofing attack model, where two malicious UAVs conduct a collusion attack while the other malicious UAVs falsify their positions.} }
 \label{F:model_mixed}
 \end{figure}

Consider a swarm of UAVs executing a routine cruising mission, during which the UAVs cooperatively maintain a specific formation to prevent collisions. Each UAV within the swarm communicates its position, ascertained by the GPS, to its counterparts. Each UAV also conducts relative distance measurements with the other UAVs that are within the permissible communication range of the UAV. Random receiver noises or GPS errors can corrupt these distance measurements, rendering the reported positions inaccurate. During the formation flight, each UAV adjusts its flight position based on the reported positions and distance measurements of neighboring UAVs, thereby averting potential flight collisions.

Malicious UAVs, under the control of attackers, have the capability to fabricate their position information and disseminate this information among all other UAVs. More precisely, malicious UAVs can initiate a spoofing attack by falsely reporting their positions within the detection measurement range of benign UAVs. This deliberate misrepresentation of positions can directly disrupt the formation. Moreover, malicious UAVs may target a specific benign UAV by deceptively reporting their positions within the detection measurement range of that UAV, which is a tactic known as a collusion attack; {\blue  see
Fig.~\ref{F:model_collusive}}. 
This coordinated misrepresentation is aimed at framing a target UAV. Given the substantial evidence presented through this deceptive conspiracy, the swarm may erroneously conclude that the framed UAV is perpetrating a spoofing attack.

The two above-mentioned attacks, i.e., the distributed attack and the collusion attack, can be amalgamated to initiate a mixed spoofing attack; {\blue  see
Fig.~\ref{F:model_mixed}}. In this composite attack, some malicious UAVs execute a distributed attack while the remaining UAVs engage in a collusion attack. The mixed nature of this attack significantly intensifies its severity, potentially hastening the breakdown of the entire formation.

\subsection{Problem Statement}
We propose to formulate the problem of identifying malicious UAVs as a localization feasibility problem. If localization is infeasible under the positions and distance measurements reported, there is at least one malicious UAV attacking the UAV swarm in an attempt to compromise the swarm formation. 

Let ${\cal X}:=\{\boldsymbol{x}_1, \ldots, \boldsymbol{x}_N\}$ define the three-dimensional (3D) coordinates of the UAVs in an $N$-UAV swarm studied, where $\boldsymbol{x}_i \in \mathbb{R}^{3 \times 1}$ is the unknown actual 3D coordinates of UAV~$i$. {\color{black} Let $\mathbb{N}=\{1,\ldots,N\}$ collect the indexes of the $N$ UAVs.} 
The permissible range for pairwise distance measurement is denoted by $d$. 
The reported position of UAV $i$ is given as $\hat{\boldsymbol{x}_{i}}$,
which can be contaminated by the measurement noise, i.e., 
$\boldsymbol{w}_{i}\sim\mathcal{N}(0,\sigma_{i}^2\boldsymbol{I}_3)$. 
$\hat{\boldsymbol{x}}_{i} = \boldsymbol{x}_{i}+\boldsymbol{w}_{i},\,\forall i$. 
Here, $\mathcal{N}(0,\sigma_{i}^2\boldsymbol{I}_3)$ stands for the zero-mean Gaussian distribution with the variance of $\sigma_{i}^2\boldsymbol{I}_3$ and $\boldsymbol{I}_3$ is the $3\times 3$ identity matrix.

The feasibility problem of finding a solution of ${\cal X}$ can be formulated as
\begin{subequations}\label{eq.iov1}
\begin{align}
&{\rm find} \ \ \cal X \notag\\ 
\text{s.t.}\ &\|\boldsymbol{x}_{i}-\hat{\boldsymbol{x}}_{j}\|^{2}< d^{2},\,
{\color{black} \forall i \in \mathbb{N}, j\in \mathbb{N}_i,}
\label{eq.prob1cons1}\\
&|\boldsymbol{x}_{i}-\hat{\boldsymbol{x}}_{i}\|^{2}\leq \epsilon,\ 
{\color{black} \forall i \in \mathbb{N},}
\label{eq.prob1cons2}  \\
&|\hat{r}_{ij}^{2}-\|\boldsymbol{x}_{i}-\hat{\boldsymbol{x}}_{j}\|^{2}|<(\frac{d}{2}{})^{2},\ 
{\color{black} \forall i \in \mathbb{N}, j\in \mathbb{N}_i,}
\label{eq.prob1cons3} 
\end{align}
\end{subequations}
where 
{\color{black} $\mathbb{N}_i$ collects all one-hop neighbors of UAV $i$ and can be obtained based on the reports of UAV $i$;}
$\epsilon\geq0 $ is a small constant; and $\hat{r}_{ij}$ denotes the reported pairwise distance between UAVs $i$ and $j$ contaminated with measurement noise,
e.g., $w_{ij}\sim\mathcal{N}(0,\sigma_{ij}^2)$. Here, $\sigma_{ij}^2$ is the variance of the distance measurement error.

Constraint (\ref{eq.prob1cons1}) defines that the relative distance measurement between UAV $i$ and the reported position of UAV $j$, i.e., $\|\boldsymbol{x}_{i}-\hat{\boldsymbol{x}}_{j}\|$, has to be within the permissible communication range, $d$, if UAV $i$ is a one-hop neighbor of UAV $j$ and can hear its ranging signals.
Constraint (\ref{eq.prob1cons2}) dictates that the difference between $\boldsymbol{x}_{i}$ and $\hat{\boldsymbol{x}_{i}}$ is smaller than a pre-specified threshold $\epsilon$, ensuring that the model must rely on the individual reported position to output a solution (if such a solution does not exist, it is reasonable to suspect there exist misreported UAV positions).
Constraint (\ref{eq.prob1cons3}) indicates that the difference between the reported pairwise distance and the Euclidean distance (between the estimated and reported positions)
should be smaller than half of the distance measurement range,
which is also considered to be the maximum tolerable distance measurement error.
Nonetheless, the right-hand side (RHS) of constraint (\ref{eq.prob1cons3}), i.e., $(\frac{d}{2})^{2}$,  can adapt to the needs of different measurement devices.

\subsection{Proposed SDP-based Reformulation}
Because of the non-convex constraints (\ref{eq.prob1cons1})--(\ref{eq.prob1cons3}), finding $\cal X$ in (\ref{eq.iov1}) is a non-convex feasibility checking problem, which is difficult to solve. 
To convexify (\ref{eq.prob1cons1})--(\ref{eq.prob1cons3}), auxiliary variables, denoted by $\alpha_{ij},\ \forall i,j \in [1,N]$, are introduced to substitute those non-convex parts in (\ref{eq.prob1cons1})--(\ref{eq.prob1cons3}).
As a result, the non-convex feasibility problem in (\ref{eq.iov1}) can be rewritten as 
\begin{subequations}\label{eq.iov1_sub}
\begin{align}
& \ \ \ \  \rm find\ \ \cal X \notag\\ 
& \text{s.t.}\ \alpha_{ij}< d^{2},\ 
{\color{black} \forall i \in \mathbb{N}, j\in \mathbb{N}_i,} \label{eq.prob1_subcons1} \\
&\ \ \ \ \alpha_{ii}\leq \epsilon,\ \forall i \in \mathbb{N},\label{eq.prob1_subcons2} \\
&\ \ \ \ |\hat{r}_{ij}^{2}-\alpha_{ij}|<(\frac{d}{2}{})^{2},\ {\color{black} \forall i \in \mathbb{N}, j\in \mathbb{N}_i,}   \label{eq.prob1_subcons3} \\
&\ \ \ \ ||\boldsymbol{x}_{i}-\hat{\boldsymbol{x}}_{j}||^{2} = \alpha_{ij},\ \ {\color{black} \forall i \in \mathbb{N}, j\in \mathbb{N}_i\cup \{i\},}\ \label{eq.prob1_subcons4} 
\end{align}
\end{subequations}
where constraints (\ref{eq.prob1_subcons1}) and (\ref{eq.prob1_subcons4}) are homogenized from (\ref{eq.prob1cons1}), and (\ref{eq.prob1_subcons2}) is homogenized from (\ref{eq.prob1cons2}). Constraints (\ref{eq.prob1_subcons1})--(\ref{eq.prob1_subcons3}) are affine and convex. 
Constraint (\ref{eq.prob1_subcons4}) is still non-convex. To convexify (\ref{eq.prob1_subcons4}), we rewrite
$||\boldsymbol{x}_{i}-\hat{\boldsymbol{x}}_{j}||^{2}$
in a matrix form as
\begin{equation}\label{eq.q1}
    \|\boldsymbol{x}_{i}-\hat{\boldsymbol{x}}_{j}\|^{2}=[\hat{\boldsymbol{x}}_{j}^{T}~-\boldsymbol{e}_{i}^{T}]\left[\begin{matrix}
       \boldsymbol{I}_{3} & \boldsymbol{X} \\
       \boldsymbol{X}^{T}   & \boldsymbol{Y} \end{matrix} \right]\left[\begin{matrix}
       \hat{\boldsymbol{x}}_{j} \\
       -\boldsymbol{e}_{i} \end{matrix} \right],
\end{equation}
where $\boldsymbol{e}_{i}\in \mathbb{R}^{N\times 1}$ is a vector whose $i$-th element is ``1'' and the rest are all ``0''. $\boldsymbol{X} \in \mathbb{R}^{3 \times N}$ is a $3 \times N$ matrix with its $i$-th column being $\boldsymbol{x}_{i}$.
Moreover, 
\begin{equation}\label{eq.y1}
\boldsymbol{Y} = \boldsymbol{X}^{T}\boldsymbol{X} \in R^{N \times N}. 
\end{equation}

As a result, finding $\cal X$ in problem (\ref{eq.iov1_sub}) can be equivalently reformulated as
\begin{subequations}\label{eq.iov2_plus}
\begin{align}
& \ \ \ \  \rm find\  \boldsymbol{X}, \boldsymbol{Y} \notag\\
&\text{s.t.}~\alpha_{ij}< d^{2},\ {\color{black} \forall i \in \mathbb{N}, j\in \mathbb{N}_i,}   \label{eq.iov2_pluscons1} \\
&\ \ \ \ \alpha_{ii}\leq \epsilon,\ \forall i \in \mathbb{N},\label{eq.iov2_pluscons2} \\
&\ \ \ \ |\hat{r}_{ij}^{2}-\alpha_{ij}|<(\frac{d}{2}{})^{2},\ {\color{black} \forall i \in \mathbb{N}, j\in \mathbb{N}_i,}   \label{eq.iov2_pluscons6} \\
&\ \ \ \ [\hat{\boldsymbol{x}}_{j}^{T}~-\boldsymbol{e}_{i}^{T}]\left[\begin{matrix}
       \boldsymbol{I}_{3} & \boldsymbol{X} \\
       \boldsymbol{X}^{T}   & \boldsymbol{Y} \end{matrix} \right]\left[\begin{matrix}
       \hat{\boldsymbol{x}}_{j} \\
       -\boldsymbol{e}_{i} \end{matrix} \right] = \alpha_{ij}, {\color{black} \forall i \in \mathbb{N}, j\in \mathbb{N}_i\cup \{i\},}\label{eq.iov2_pluscons3} \\
&\ \ \ \ \boldsymbol{Y} = \boldsymbol{X}^{T}\boldsymbol{X}. \label{eq.iov2_pluscons4} 
\end{align}
\end{subequations}
Here, constraints (\ref{eq.iov2_pluscons3}) and (\ref{eq.iov2_pluscons4}) are non-convex. Yet, constraint (\ref{eq.iov2_pluscons4}) can be relaxed as~\cite{bi_iot2023,Biswas06}
\begin{equation}\label{eq.yy1}
\boldsymbol{Y} \succeq \boldsymbol{X}^{T}\boldsymbol{X}, 
\end{equation}
{\color{black} where ``$\succeq$'' stands for element-wise inequality.}
According to Schur complement \cite{boyd_vandenberghe_2004}, (\ref{eq.yy1}) is equivalent to 
\begin{equation}\label{eq.yyy1}
\left[\begin{matrix}
       \boldsymbol{I}_{3} & \boldsymbol{X} \\
       \boldsymbol{X}^{T}   & \boldsymbol{Y} \end{matrix} \right] \succeq 0,
\end{equation}
Further let $\boldsymbol{{\cal{Z}}}$ denote the left-hand side (LHS) of (\ref{eq.yyy1}), yielding
\begin{equation}\label{eq.q3}
    \boldsymbol{{\cal{Z}}}=\left[\begin{matrix}
       \boldsymbol{I}_{3} & \boldsymbol{X} \\
       \boldsymbol{X}^{T}   & \boldsymbol{Y} \end{matrix} \right] \succeq 0.
\end{equation}
The relaxation of (\ref{eq.iov2_pluscons4}) to \eqref{eq.yy1} is tight if
${\rm Rank}(\boldsymbol{{\cal{Z}}})=3$.

Also, define
\begin{equation}\label{eq.q2}
    \boldsymbol{\hat{{\cal G}}_{ij}}=\left[\begin{matrix}
       \hat{\boldsymbol{x}}_{j} \\
       -\boldsymbol{e}_{i} \end{matrix} \right][\hat{\boldsymbol{x}}_{j}^{T}~-\boldsymbol{e}_{i}^{T}].
\end{equation} 
Based on (\ref{eq.q3}) and (\ref{eq.q2}), (\ref{eq.iov2_pluscons3}) can be rewritten as  
 \begin{equation}\label{eq.t1}
  [\hat{\boldsymbol{x}}_{j}^{T}~-\boldsymbol{e}_{i}^{T}]\left[\begin{matrix}
       \boldsymbol{I}_{3} & \boldsymbol{X} \\
       \boldsymbol{X}^{T}   & \boldsymbol{Y} \end{matrix} \right]\left[\begin{matrix}
       \hat{\boldsymbol{x}}_{j} \\
       -\boldsymbol{e}_{i} \end{matrix} \right]=\text{Tr}(\boldsymbol{\hat{{\cal G}}_{ij}}\boldsymbol{{\cal{Z}}}) = \alpha_{ij},\ \ \\
 \end{equation}
Based on (\ref{eq.yy1})--(\ref{eq.t1}), the feasibility problem (\ref{eq.iov2_plus}) is further equivalently rewritten as the following feasibility problem: 
\begin{subequations}\label{eq.iov3}
\begin{align}
& \ \ \ \  \rm find\  \boldsymbol{{\cal{Z}}} \notag\\
&\text{s.t.}~\boldsymbol{{\cal{Z}}}_{1:3,1:3}=\boldsymbol{I}_{3},\label{eq.prob3cons0}\\
&\ \ \ \ \alpha_{ij}< d^{2},\ {\color{black} \forall i \in \mathbb{N}, j\in \mathbb{N}_i,}   \label{eq.prob3cons1} \\
&\ \ \ \ \alpha_{ii}\leq \epsilon,\ \forall i \in \mathbb{N},\label{eq.prob3cons2} \\
&\ \ \ \ |\hat{r}_{ij}^{2}-\alpha_{ij}|<(\frac{d}{2}{})^{2},\ {\color{black} \forall i \in \mathbb{N}, j\in \mathbb{N}_i,}   \label{eq.prob3cons6} \\
&\ \ \ \ \text{Tr}(\boldsymbol{\hat{{\cal G}}_{ij}}\boldsymbol{{\cal{Z}}}) = \alpha_{ij},\ \ {\color{black} \forall i \in \mathbb{N}, j\in \mathbb{N}_i\cup \{i\},}\label{eq.prob3cons3} \\
&\ \ \ \ \boldsymbol{{\cal{Z}}} \succeq 0, \label{eq.prob3cons4} \\
&\ \ \ \ \text{Rank}(\boldsymbol{{\cal{Z}}})=3. \label{eq.prob3cons5}
\end{align}
\end{subequations}
where constraint (\ref{eq.prob3cons0}) enforces the upper left 3$\times$3 block of $\boldsymbol{{\cal{Z}}}$ to be an identity matrix, ensuring that the rank of the solution is at least three. Constraints (\ref{eq.prob3cons0}), 
(\ref{eq.prob3cons4}), and (\ref{eq.prob3cons5}) are equivalently derived from (\ref{eq.iov2_pluscons4}). This is because both (\ref{eq.prob3cons0}) and 
(\ref{eq.prob3cons4}) constrain $\boldsymbol{{\cal{Z}}}$ to be symmetric and in the form of (\ref{eq.q3}), while rank constraint (\ref{eq.prob3cons5}) forces the lower right $N \times N$ block of $\boldsymbol{{\cal{Z}}}$, i.e., $\boldsymbol{Y}$ in (\ref{eq.q3}), to be $\boldsymbol{X}^{T}\boldsymbol{X}$, according to classic linear algebra theory.

Dropping the rank constraint (\ref{eq.prob3cons5}), we have the following SDR problem:
\begin{subequations}\label{eq.iov4}
\begin{align}
& \ \ \ \  \rm find\  \boldsymbol{{\cal{Z}}} \notag\\
&\text{s.t.}~\boldsymbol{{\cal{Z}}}_{1:3,1:3}=\boldsymbol{I}_{3},\label{eq.prob4cons0}\\
& \ \ \ \ \alpha_{ij}< d^{2},\ {\color{black} \forall i \in \mathbb{N}, j\in \mathbb{N}_i,}   \label{eq.prob4cons1} \\
&\ \ \ \ \alpha_{ii}\leq \epsilon,\ \forall i \in \mathbb{N},\label{eq.prob4cons2} \\
&\ \ \ \ |\hat{r}_{ij}^{2}-\alpha_{ij}|<(\frac{d}{2}{})^{2},\ {\color{black} \forall i \in \mathbb{N}, j\in \mathbb{N}_i,}   \label{eq.prob4cons5} \\
&\ \ \ \ \text{Tr}(\boldsymbol{\hat{{\cal G}}_{ij}}\boldsymbol{{\cal{Z}}}) = \alpha_{ij},\ \ {\color{black} \forall i \in \mathbb{N}, j\in \mathbb{N}_i\cup \{i\},}\label{eq.prob4cons3} \\
&\ \ \ \ \boldsymbol{{\cal{Z}}} \succeq 0. \label{eq.prob4cons4}
\end{align}
\end{subequations}

Problem (\ref{eq.iov4}) is convex and can be efficiently solved using off-the-shelf CVX solvers, e.g., MATLAB CVX toolbox~\cite{2008CVX}.
Clearly, problem (\ref{eq.iov4}) is a relaxed (but generally tight) version of the original feasibility problem (\ref{eq.iov1}), with a larger feasible solution region. If the problem in (\ref{eq.iov4}) is infeasible, i.e., no feasible solution exists for the problem in (\ref{eq.iov4}), 
then problem (\ref{eq.iov1}) is surely infeasible.
As a result, we can detect whether at least one malicious UAV misreports its position by checking the feasibility of the problem in~(\ref{eq.iov4}). 

\section{Proposed approach for malicious UAV identification}
\label{approach}
Leveraging the SDP problem in~(\ref{eq.iov4}), we proceed to develop two new algorithms, CDI and E-CDI, to exploit the proximity of adjacent UAVs to facilitate the effective detection of malicious UAVs and eliminate spoofing attacks. 

\subsection{Initialization of Malicious UAV Set}\label{sec: initial malicious set determination}
{\blue

Let $\mathbb{M}$ and $\mathbb{B}$ denote the sets of malicious and benign UAVs, respectively. $\mathbb{M} \cup \mathbb{B} = \mathbb{N}$. Based on $\boldsymbol{E}_{r}$ and $\boldsymbol{E}_{n}$, we propose to initialize $\mathbb{M}$ and $\mathbb{B}$, as follows.

We come up with two Euclidean matrices, i.e., the generated Euclidean matrix from individual reported positions, denoted by $\boldsymbol{E}_{r}\in R^{N\times N}$, and the detected Euclidean distances matrix contaminated with noise, denoted by $\boldsymbol{E}_{n}\in R^{N\times N}$, as
\begin{equation}\label{eq.er1}
\boldsymbol{E}_{r} =\begin{pmatrix}
    \rho_{11}\|\hat{\boldsymbol{x}_{1}}-\hat{\boldsymbol{x}_{1}}\| &  \cdots     &     \rho_{1N}\|\hat{\boldsymbol{x}_{1}}-\hat{\boldsymbol{x}_{N}}\|  \\
\vdots  & \ddots     & \vdots        \\
    \rho_{N1}\|\hat{\boldsymbol{x}_{N}}-\hat{\boldsymbol{x}_{1}}\| & \cdots      &    \rho_{NN}\|\hat{\boldsymbol{x}_{N}}-\hat{\boldsymbol{x}_{N}}\|   \\
\end{pmatrix},
\end{equation}
\begin{equation}\label{eq.en1}
\boldsymbol{E}_{n} =\begin{pmatrix}
\hat{r}_{11} &  \cdots     &     \hat{r}_{1N} \\
\vdots  & \ddots     & \vdots        \\
\hat{r}_{N1} & \cdots      &    \hat{r}_{NN}   \\
\end{pmatrix},
\end{equation}
where  $\rho_{ij}$ indicates if UAVs $i$ and $j$ are directly connected. $\rho_{ij}=1$, if UAVs $i$ and $j$ are within each other's permissible communication range, i.e., $\hat{r}_{ij}>0$; otherwise, $\rho_{ij}=0$.
In this sense, $\boldsymbol{E}_{r}$ can be a sparse matrix (like $\boldsymbol{E}_{n}$), depending on the communication range of the UAV.

We can carry out element-wise comparisons between $\boldsymbol{E}_{r}$ and $\boldsymbol{E}_{n}$. Specifically, if the $(i,j)$-th elements of the two matrices have a smaller difference than the pre-specified threshold $\frac{d}{2}$, i.e., \eqref{eq.prob1cons3} is unsatisfied, then UAVs $i$ and/or $j$ are potentially malicious. Both of the UAVs are added into $\mathbb{M}$, i.e., $\mathbb{M}=\mathbb{M} \cup \{i,j\}$. After all $N^2$ elements of $\boldsymbol{E}_{r}$ and $\boldsymbol{E}_{n}$ are assessed, the initial $\mathbb{M}$ is obtained. $\mathbb{B}$ can be accordingly initialized to be $\mathbb{B} =\mathbb{N} \setminus \mathbb{M}$.

 }

\subsection{
Proposed CDI Algorithm}

\begin{algorithm}[t]
\caption{The proposed  CDI algorithm}\label{algorithm}
\KwIn{$\mathbb{B}$; $\mathbb{M}$; $\hat{\boldsymbol{x}_{i}}$, $\forall i \in [1,N]$;  $\hat{r_{ij}}$,\ $\forall i,j \in [1,N],\ i\neq j$; $d$; $\epsilon$.}
\KwOut{$\mathbb{M}$}
Set $k\gets 1$;\\
 \While{$\mathbb{M}~ \textit{can be further reduced}$}
{
Select the $k$-th UAV of $\mathbb{M}$ and the set of its one-hop neighbors $\mathbb{N}_k$;\\
Construct $\mathbb{T}\gets \mathbb{B}\cup\{k,\mathbb{N}_k\}$;\\
Apply $\mathbb{T}$ to (\ref{eq.iov4}), and check feasibility using SDP.

\If{problem (\ref{eq.iov4}) is feasible $\mathbb{T}$}
{
$\mathbb{M}\gets\mathbb{M}\setminus \{k,\mathbb{N}_k \}$;\\
$\mathbb{B} \gets \mathbb{B}\cup \{k, \mathbb{N}_k\}$;
}
Set $k\gets(k+1)\text{ mod }|\mathbb{M}|$;

}

\end{algorithm}

As summarized in \textbf{Algorithm 1}, we propose to assess the potentially malicious UAVs in $\mathbb{M}$ one after another and move those actually benign from $\mathbb{M}$ to $\mathbb{B}$ until both $\mathbb{M}$ and $\mathbb{B}$ stop changing. When assessing a UAV, i.e., UAV $k$, from $\mathbb{M}$, we also consider its one-hop neighbors. 
Part of $\mathbb{N}_k$ may belong to $\mathbb{B}$, and the rest belong to $\mathbb{M}$.

We apply the feasibility checking problem in (12) to the collection of $\mathbb{B}$, $\{k\}$, and $\mathbb{N}_k$, i.e., $\mathbb{B}\cup \{k, \mathbb{N}_k\}$. If the problem is feasible, UAV $k$ and its one-hop neighbors $\mathbb{N}_k$ are benign. They can be removed from $\mathbb{M}$ and added to $\mathbb{B}$; i.e., $\mathbb{M}=\mathbb{M}\setminus \{k,\mathbb{N}_k \}$ and $\mathbb{B}=\mathbb{B}\cup \{k,\mathbb{N}_k \}$. Otherwise, they remain in~$\mathbb{M}$. 
The reason for considering a potentially malicious UAV $k$ together with its one-hop neighbors $\mathbb{N}_k$ is to increase the chance that UAV $k$ is connected to $\mathbb{B}$. Therefore, the feasibility checking problem can be meaningfully carried out. In the case where UAV $k$ and its one-hop neighbors $\mathbb{N}_k$ are disconnected from $\mathbb{B}$ (in other words, $\mathbb{N}_k$ and their neighbors all belong to $\mathbb{M}$), then UAV $k$ and its one-hop neighbors $\mathbb{N}_k$ remain in $\mathbb{M}$. 

In this way, we repeatedly assess the remaining UAVs in $\mathbb{M}$ until $\mathbb{M}$ cannot be further reduced. This algorithm can quickly detect and identify malicious UAVs; but may overkill, i.e., misjudge a benign UAV to be malicious in the situation where the benign UAV only has a malicious one-hop neighbor since they are always assessed together for feasibility and cannot be individually arbitrated. In this sense, the algorithm is conservative and can be overprotective. 


\subsection{
Proposed E-CDI Algorithm}\label{sec: e-CDI}

\begin{figure*} \center{\includegraphics[width=15 cm]  {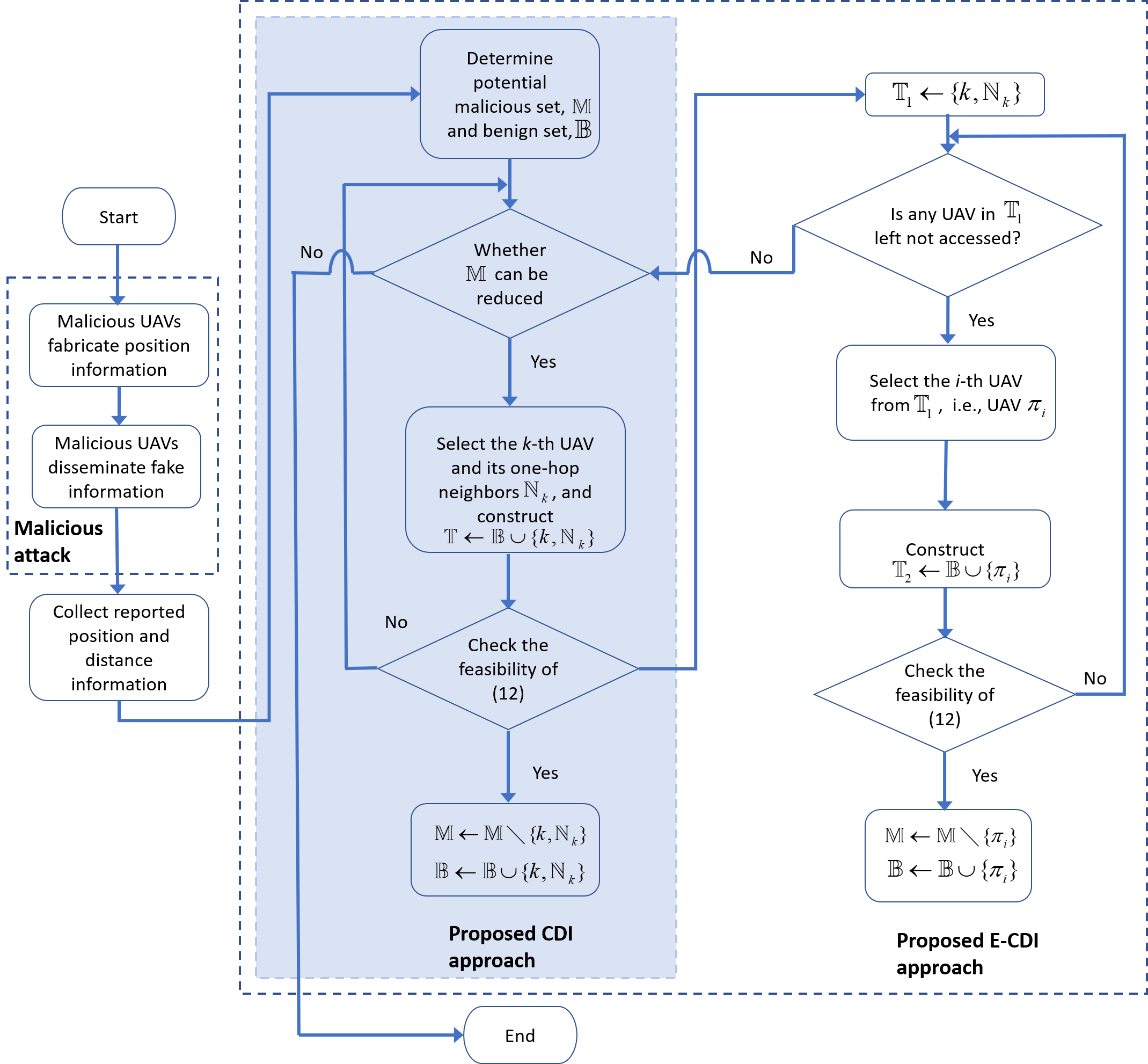}}
 \caption{\label{1} {\blue The flowchart of the proposed CDI/E-CDI algorithm, where the shaded part accounts for the CDI algorithm, which is part of the more comprehensive E-CDI algorithm.} }
 \label{F:flowchart}
 \end{figure*}

A key difference between the E-CDI algorithm and the CDI algorithm (\textbf{Algorithm 1}) is that the E-CDI algorithm assesses each of the potentially malicious one-hop neighbors of a UAV belonging to $\mathbb{M}$ individually, each time the UAV and its one-hop neighbors fail the feasibility check. Specifically, each of UAV $k$ and its potentially malicious one-hop neighbors in $\mathbb{N}_k$ are assessed by temporarily joining $\mathbb{B}$ for feasibility check again. Those that turn out to be benign are removed from $\mathbb{M}$ and added to $\mathbb{B}$. By this means, each connected malicious UAV can be detected and identified. The details are provided in \textbf{Algorithm~2}. The flowchart of the proposed CDI/E-CDI algorithm is provided in Fig.~\ref{F:flowchart}. 

Another key difference is that the E-CDI algorithm is able to detect collusion attacks, while the CDI algorithm cannot. This is because the E-CDI assesses individual UAVs in a neighborhood $\{k,\mathbb{N}_k\}$ if the neighborhood is detected to be infected by malicious UAVs in the neighborhood. As a result, the malicious UAVs (or UAVs that cannot be confirmed benign due to their poor connectivity to other benign UAVs) can be individually assessed and verified. In contrast, the CDI algorithm may not achieve this since its assessment is based on neighborhoods~$\{k,\mathbb{N}_k\},\, \forall k \in \mathbb{M}$.



\begin{algorithm}[t]
\caption{The proposed E-CDI algorithm}\label{algorithm2}
\KwIn{$\mathbb{B}$; $\mathbb{M}$; $\hat{\boldsymbol{x}_{i}}$, $\forall i \in [1,N]$;  $\hat{r_{ij}}$,\ $\forall i,j \in [1,N],\ i\neq j$; $d$; $\epsilon$.}
\KwOut{$\mathbb{M}$}
Set $k\gets 1$; \\
 \While{$\mathbb{M}~ \textit{can be further reduced}$}
{
Select the $k$-th UAV in $\mathbb{M}$ and the set of its one-hop neighbors $\mathbb{N}_k$;\\
Construct $\mathbb{T}\gets \mathbb{B}\cup\{k,\mathbb{N}_k\}$;\\
Apply $\mathbb{T}$ to (\ref{eq.iov4}), and check feasibility using SDP.\\
\eIf{problem (\ref{eq.iov4}) is feasible upon $\mathbb{T}$}
{
$\mathbb{M}\gets\mathbb{M}\setminus \{k,\mathbb{N}_k \}$;\\
$\mathbb{B} \gets \mathbb{B}\cup \{k, \mathbb{N}_k\}$;
}
{
Set $\mathbb{T}_1 \gets \{k,\mathbb{N}_k\}$; $i \gets 1$; \\
\While{i $\le$ $|\mathbb{T}_1|$ }
{
  Select the $i$-th UAV of $\{k,\mathbb{N}_k\}$, denoted by~$\pi_i$;\\
  Construct $\mathbb{T}_2\gets\mathbb{B}\cup\{\pi_i\}$;\\
  
  Apply $\mathbb{T}_2$ to (\ref{eq.iov4}) and check feasibility using SDP;\\
  \If{ problem (\ref{eq.iov4}) is feasible upon $\mathbb{T}_2$}
  {
  $\mathbb{M}\gets\mathbb{M}\setminus \{\pi_{i} \}$;\\
$\mathbb{B} \gets \mathbb{B}\cup \{\pi_{i}\}$;\\   
  }
  $i \gets i+1$;\\
}

}
Set $k\gets(k+1)\text{ mod }|\mathbb{M}|$;
}

\end{algorithm}

\section{Simulation Results}

\begin{table}
\centering 
\caption{ Simulation parameters and configuration.}
\label{table_settings}
\begin{tabular}{p{3.5cm}  p{3cm}} 
\hline
 \textbf{Parameter} & \textbf{Configuration}   \\ \hline 

  Simulation environment & Matlab R2020b
   \\ 
Solver & CVX solver   \\ 
UAV swarm range & Unit cube $[-0.5,+0.5]^{3}$   \\ 
Reported position noise & $\boldsymbol{w}_{i}\sim\cal{N}$$(0,10^{-6}\boldsymbol{I}_3)$   \\ 
Distance measurement noise & $w_{ij}\sim\cal{N}$$(0,10^{-6})$  \\ 
Distance measurement range & $[0.25,0.45]^{3}$  \\ 
     \hline

\end{tabular}
\end{table}

In this section, we consider three types of spoofing attacks to gauge the capability of the proposed algorithm to counteract these attacks. 
We conduct extensive simulations to comprehensively evaluate the proposed algorithms in comparison with the established benchmarks on the key factors, i.e., the number of malicious UAVs, the scale of the network, distance measurement noise, and measurement distance.

\subsection{
Simulation Setting
}
We consider a UAV swarm with the UAVs' positions randomly generated according to a uniform distribution inside a unit cube $[-0.5,+0.5]^{3}$. The malicious UAVs are randomly chosen from the nodes. The distance measurement range is $d=0.3$. Note that both the reported positions and reported distance measurements are contaminated with additive Gaussian noises, $\boldsymbol{w}_{i}\sim\cal{N}$$(0,10^{-6}\boldsymbol{I}_3)$~\cite{location_accuracy} 
 and $w_{ij}\sim\cal{N}$$(0,10^{-6})$, respectively.
 {\blue The key parameters of the simulations are summarized in Tab.~\ref{table_settings}.}

We assess the performances of the proposed algorithms against three types of position spoofing attacks, as follows.
\begin{itemize}
    \item \textbf{Distributed spoofing attack}
    Under this attack, several malicious UAVs independently misreport their positions in an attempt to compromise the UAV swarm formation.
    \item \textbf{Collusion attack}
    Under this attack, several malicious UAVs conspire to frame some benign UAVs and make them falsely identified as malicious.
    Based on the reported positions from targeted benign UAVs,
    the malicious UAVs misreport their positions to be within the neighborhood of those benign UAVs, though they can be far away from the UAVs under attack.

    
    \item \textbf{Mixed spoofing attack}
    Under this attack, malicious UAVs launch attacks in both distributed and collusive fashions.
    Specifically, some of the malicious UAVs independently carry out distributed spoofing attacks to compromise the swarm formation.
    The rest of the malicious UAVs cooperate to further impair or corrode the integrity of the UAV swarm.  
\end{itemize}


\begin{figure}[t]
\centering \epsfig{file=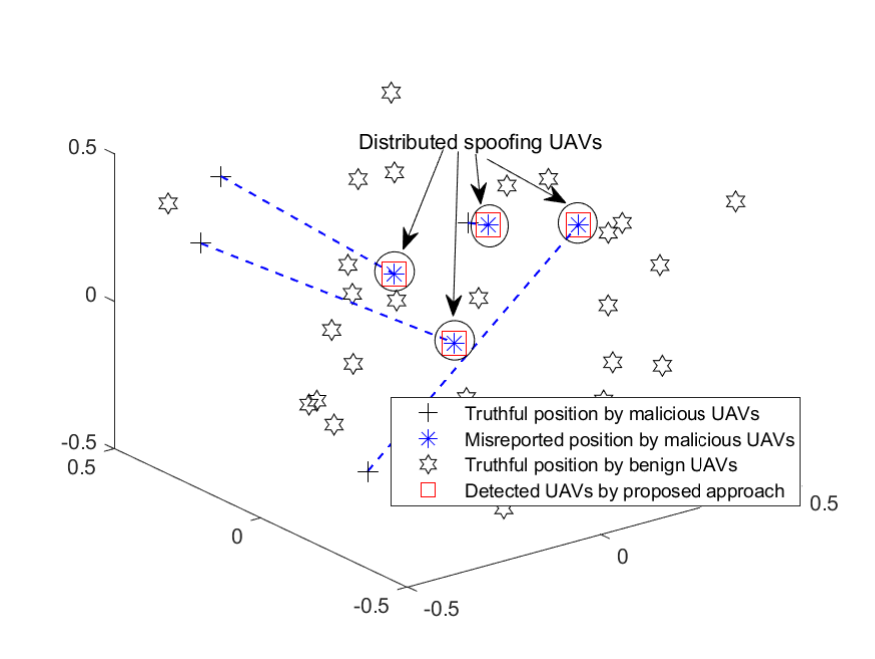, width=0.53\textwidth} 
\caption{ An identification sample of the proposed approach against distributed spoofing attack in a UAV swarm with $N=30$ UAVs and $M=4$ malicious UAVs, where $d=0.3$. The dashed lines exhibit the misreported distance.}
\label{F:UAV1} 
\end{figure}

\begin{figure}[t]
\centering \epsfig{file=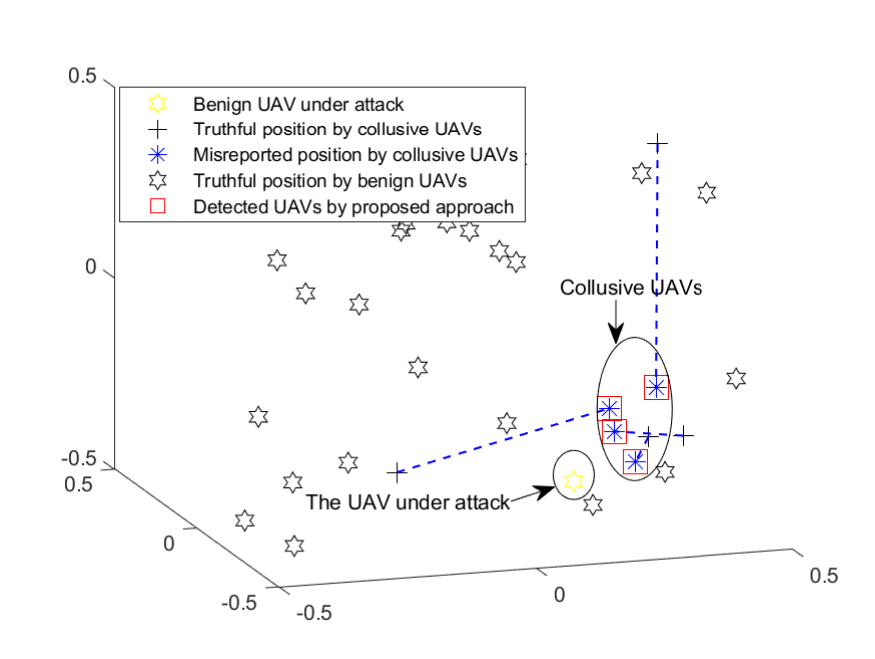, width=0.53\textwidth} 
\caption{ An identification sample of the proposed approach against collusion spoofing attack, in a UAV swarm with $N=30$ UAVs and $M=4$ malicious UAVs, where $d=0.3$. The dashed lines exhibit the misreported distance.}
\label{F:UAV2} 
\end{figure}

\begin{figure}[t]
\centering \epsfig{file=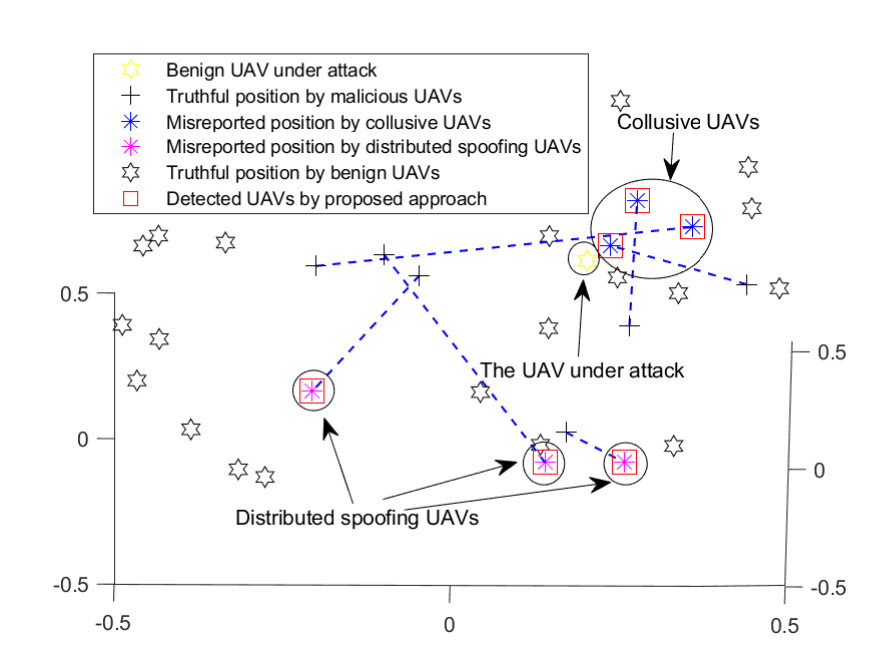, width=0.53\textwidth} 
\caption{ An identification sample of the proposed approach against mixed spoofing attack, in a UAV swarm with $N=30$ UAVs and $M=6$ malicious UAVs, where $d=0.3$. The dashed lines exhibit the misreported distance.}
\label{F:UAV3} 
\end{figure}

The benchmark algorithms considered are
\begin{itemize}
 
\item \textbf{NLOS-based approach:}
This approach \cite{NLOS} treats errors induced by the misbehavior of malicious UAVs as a variant of NLOS, since NLOS and spoofed positions are alike, i.e., causing considerable deviations from the genuine positions. However, NLOS is a path error involving two UAVs in a swarm.
Therefore, we mildly adjust it to suit comparison by random sampling according to the scale of the test sample. 

\item \textbf{Random approach:}
This approach directly relies on the earlier potential candidate set of malicious UAVs to conduct random sampling adjusted to the number of malicious UAVs and the sampled UAVs as the output of the approach. 
\end{itemize}

The performance metrics considered are Precision, Recall, and F1. 
The three classic metrics are given by \cite{bisus22}
\begin{subequations}\label{eq.precall}
\begin{align}
&{\rm Precision} = \frac{|Q_{p}\cap Q_{t}|}{|Q_{p}|},  \label{eq.precison} \\
&{\rm Recall} = \frac{|Q_{p}\cap Q_{t}|}{|Q_{t}|},  \label{eq.recall} \\ 
&{\rm F1} = \frac{2\times {\rm Precision} \times {\rm Recall}}{{\rm Precision+Recall}}, \label{eq.f1} 
\end{align}
\end{subequations}
where $Q_p$ stands for the predicted set by a specific algorithm, $Q_t$ denotes the ground-truth test set, and $|\cdot|$ denotes cardinality. 

To evaluate the effect of the network topology,
we consider two other metrics, including ``malicious ratio'', i.e., 
the ratio of 
the number of malicious UAVs detected initially (as done in Section~\ref{sec: initial malicious set determination}) to the total number of UAVs, denoted by $R_M = |\mathbb{M}|/|\mathbb{N}|$. 
Correspondingly, the ``benign ratio'', i.e., the ratio of the number of UAVs initially determined as benign to the total number of UAVs, is~$R_B=1-R_M$. 

\subsection{Visualization of Proposed Malicious UAV Detection}
 
Fig. \ref{F:UAV1} shows the identifications of malicious UAVs in a UAV swarm with $N=30$ UAVs under distributed spoofing attacks. There are four malicious UAVs launching distributed spoofing attacks. Both the proposed CDI and E-CDI algorithms are simulated.
Fig. \ref{F:UAV2} shows the identifications of malicious UAVs in the 30-UAV swarm under collusion attacks, where four malicious UAVs conspire collusion attacks 
towards a benign UAV. 
In Fig. \ref{F:UAV2}, we only simulate the proposed E-CDI algorithm 
since the CDI algorithm is inapplicable to collusion attacks, as discussed in Section~\ref{sec: e-CDI}.
Fig. \ref{F:UAV3} shows the identifications of malicious UAVs in the 30-UAV swarm under mixed attacks, where three malicious UAVs conspire collusion attacks 
towards a benign UAV, while three other malicious UAVs launch distributed attacks. 
We only run the proposed E-CDI algorithm in Fig. \ref{F:UAV3}
for the same reason as considered for Fig.~\ref{F:UAV2}. 
From Figs. \ref{F:UAV1}, \ref{F:UAV2}, and \ref{F:UAV3}, we can see that the proposed CDI and E-CDI algorithms can effectively detect and identify malicious UAVs in applicable scenarios.

\subsection{
Resistance to Distributed Spoofing Attacks}
\label{distributed}

\begin{figure}[t]
\centering \epsfig{file=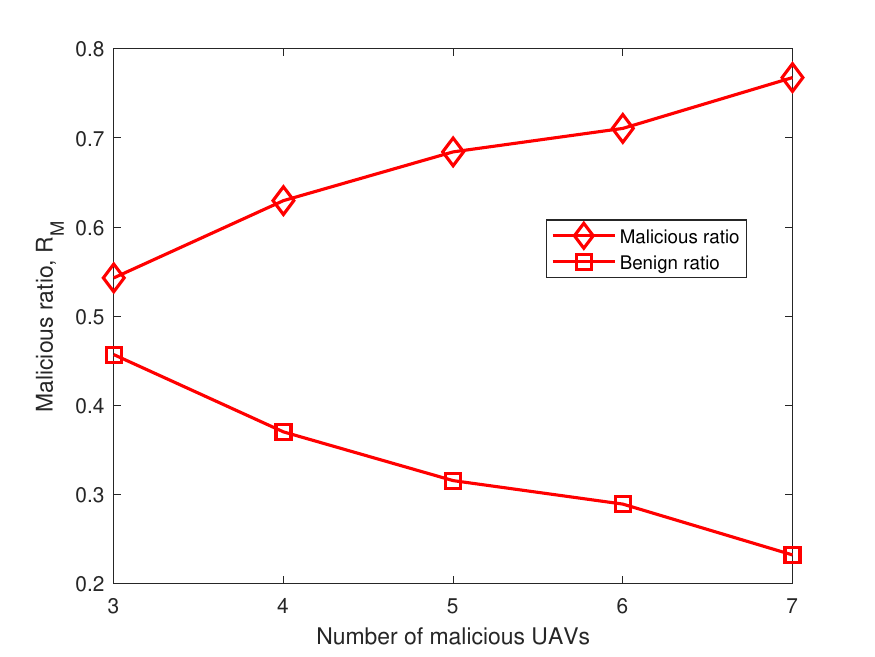, width=0.5\textwidth} 
\caption{ The variations of malicious ratio.}
\label{F:UAV4} 
\end{figure}

Fig. \ref{F:UAV4} plots the malicious ratio and average available degree across different numbers of malicious UAVs, where the average of 100 independently randomly generated swarms with consistent parameters with Fig. \ref{F:UAV1} is plotted. 
It is noticed that the malicious ratio is nearly linear to the number of malicious UAVs. When there are seven malicious UAVs, the potential malicious ratio can reach zero, highlighting the presence of multiple malicious UAVs
can severely disrupt or even dismantle the normal operation of a swarm.

\begin{figure*}[t]
\centering \epsfig{file=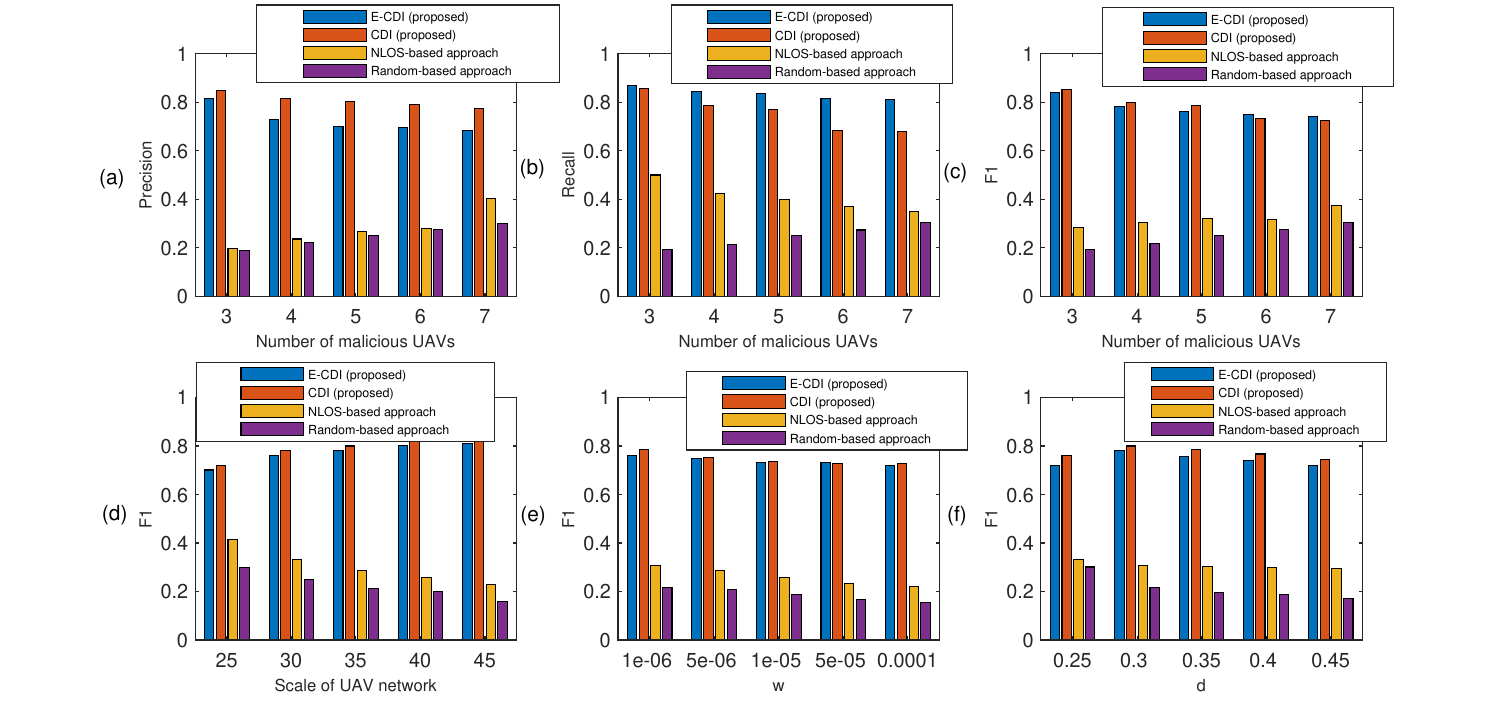, width=1.05\textwidth} 
\caption{ (a) The performance on Precision of the proposed and baseline approaches. (b) The performance on Recall of the proposed and baseline approaches. (c) The performance on F1 of the proposed and baseline approaches. (d) The performance on F1 of the proposed and baseline approaches under different scales of the swarm network. (e) The performance on F1 of the proposed and baseline approaches under different levels of distance measurement noise. (f) The performance on F1 of the proposed and baseline approaches under different levels of measurement distance.}
\label{F:UAV_combine} 
\end{figure*}

In Fig. \ref{F:UAV_combine}(a), we observe the Precision of the proposed algorithms compared to the benchmark methods. It is evident that both the CDI and E-CDI algorithms outperform the others significantly. The difference in Precision between the two algorithms can be attributed to the strategy they employ for handling neighboring index sets. The E-CDI algorithm, which selects multiple indices simultaneously, may introduce some redundancies, leading to a slightly inferior Precision compared to the CDI algorithm, which selects one index at a time. Additionally, the decreasing trend in performance of both proposed approaches can be linked to the degradation of the network structure, as indicated by the increasing malicious ratio in Fig. \ref{F:UAV4}.

On the other hand, there is a noticeable upward trend in the detection of malicious UAVs under the NLOS-based approach, especially those with significant distance errors akin to NLOS conditions. 
The NLOS-based approach, which involves selecting UAVs with the largest distance errors, simultaneously amplifies the likelihood of encountering malicious UAVs. 
The ascending trend in the Random approach can be explained by the situation where the rate of capturing malicious UAVs surpasses the expansion rate of the malicious set.

In Fig. \ref{F:UAV_combine}(b), we examine the Recall of the proposed methods compared to the benchmarks across varying numbers of malicious UAVs. Both the CDI and E-CDI algorithms outperform all benchmarks significantly. Moreover, E-CDI offers higher Recall than CDI. The reason is that E-CDI sacrifices some precision, potentially introducing redundancies, to ensure the capture of more malicious UAVs. 
Given the critical security nature of the problem studied, it is imperative to emphasize high Recall to identify as many malicious UAVs as possible to safeguard the swarm. 
As also noticed, the Recall of both proposed algorithms declines due to the deteriorating network structure, as discussed in Fig. \ref{F:UAV_combine}(a). For the same reason, the Recall of the NLOS-based approach also declines, as discussed in Fig. \ref{F:UAV_combine}(a). As for the Random-based approach, it is intriguing to note that the trends in Precision in Fig. \ref{F:UAV_combine}(a) and Recall in Fig.~\ref{F:UAV_combine}(b) bear resemblance. The conclusion drawn is that the expansion rate of captured malicious UAVs exceeds that of the malicious set enlargement, leading to this consistent trend.

Fig. \ref{F:UAV_combine}(c) illustrates the F1 of the proposed methods alongside the benchmarks across varying numbers of malicious UAVs, where the proposed CDI and E-CDI algorithms consistently outperform all other algorithms.
It is noticed that the E-CDI algorithm is initially better than the CDI algorithm. However, as the number of malicious UAVs increases, the CDI algorithm gradually surpasses the E-CDI algorithm when the number of malicious UAVs exceeds five. 
The reason is that when there are only a small number of malicious UAVs in a swarm, the E-CDI algorithm is more likely to misclassify benign UAVs as malicious. 
As the number of malicious UAVs increases and the network structure deteriorates, the E-CDI algorithm is increasingly advantageous. With a higher number of malicious UAVs, the E-CDI algorithm exhibits a greater likelihood of correctly detecting malicious UAVs.

We also notice in Fig. \ref{F:UAV_combine}(c) that the NLOS-based approach outperforms the Random approach due to its selection from a relatively smaller malicious set with a higher probability, achieved through sorting based on absolute distance error. In contrast, the Random approach selects from a larger set encompassing all possible malicious UAVs. The increasing trend observed in both benchmark algorithms is attributed to the growing number of malicious UAVs, resulting in a higher probability of detection by both benchmarks.


Fig. \ref{F:UAV_combine}(d) presents the F1 performance of our proposed algorithms compared to the benchmark algorithms across these varying network scales. Both the proposed CDI and E-CDI algorithms consistently outperform the other algorithms.
An intriguing observation is that the CDI algorithm surpasses the E-CDI algorithm across all network scales. This is primarily attributed to the fixed number of malicious UAVs: As the network scales up, a larger number of benign UAVs are likely to be present, offering supporting evidence. However, this also introduces a risk for the E-CDI algorithm, which, while aiming for higher Recall, may inadvertently incorporate benign UAVs. By contrast, the CDI algorithm, which emphasizes higher Precision by meticulously identifying one UAV at a time, mitigates this risk.
 On the other hand, the declining trend observed in both benchmarks can be attributed to the larger network scale, leading to a lower probability of detection for both benchmarks. In particular, the increased number of malicious UAVs in a larger-scale network makes it hard for the benchmarks, which employ methods like selecting the largest absolute distance error or sampling from the potential malicious set, to identify the malicious UAVs effectively.

Fig. \ref{F:UAV_combine}(e) presents the F1 performance of the proposed algorithms in comparison to the benchmarks across different levels of distance measurement inaccuracies. 
It is evident that all the algorithms exhibit a consistent declining trend, resulting from the adverse influence of inaccurate distance measurements on the effectiveness of constraints in (\ref{eq.iov4}). Nevertheless, the proposed algorithms maintain efficiency and robustness even in the face of elevated levels of measurement noise. This resilience stems from the fact that the algorithms leverage the entire swarm to aggregate evidence, effectively mitigating the impact of inaccurate distance measurements.
 On the other hand, the benchmarks face challenges with an increase in noise, as these inaccuracies can lead to greater disparities between the distances computed based on reported positions and the reported distance measurements. This, in turn, expands the candidate set of malicious UAVs, reducing the likelihood of detection by the benchmarks.


Fig. \ref{F:UAV_combine}(f) provides an overview of the F1 performance of the proposed algorithms and the benchmarks across different measurement distances. In the case of the proposed algorithms, both initially exhibit an ascending trend, followed by a descent after reaching a threshold of 0.35. This is because the increase in measurement distance prompts more UAVs to become neighbors, thereby generating additional evidence for decision-making at the beginning. However, the larger measurement distance relaxes the constraints applied to directly counter-spoofing, permitting more malicious UAVs to evade detection and the F1 declines. 
In contrast, when considering the benchmark algorithms, the expansion of measurement distance implies that more neighbors are involved in the evidence-gathering process. This can result in a higher number of UAVs reporting pairwise distance measurements, consequently leading to greater disparities between the distances computed based on reported positions and the reported distance measurements. 
The resulting larger candidate set of malicious UAVs reduces the probability of detection by the benchmarks.



\begin{figure}[t]
\centering \epsfig{file=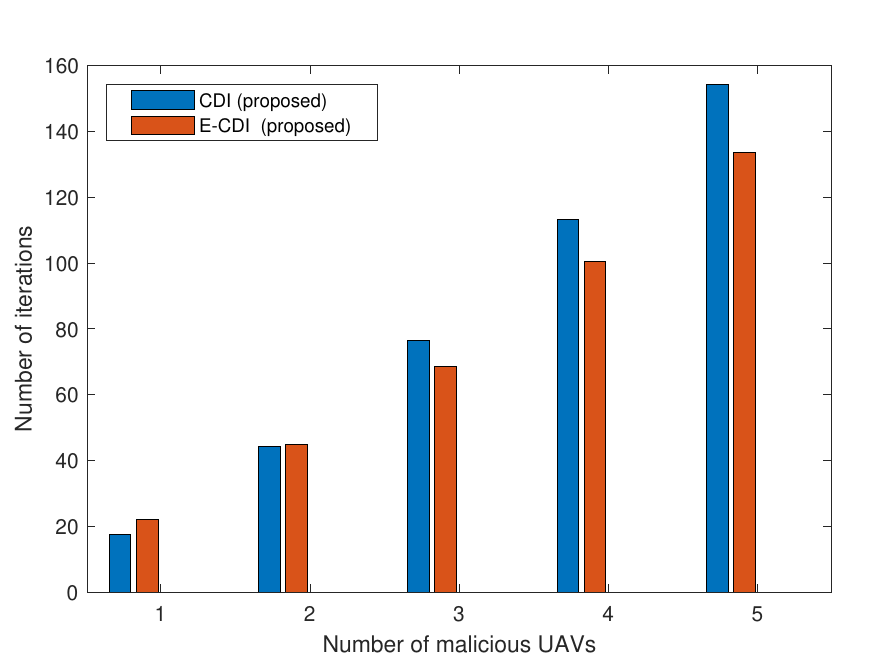, width=0.45\textwidth} 
\caption{ The number of iterations required by the CDI and E-CDI algorithms.}
\label{F:UAV_iteration} 
\end{figure}

We further compare the proposed CDI and E-CDI algorithms in the number of iterations, as shown in Fig. \ref{F:UAV_iteration}. All parameters are kept consistent with those in Fig. \ref{F:UAV1}, except for the number of malicious UAVs.
When the number of malicious UAVs is relatively small, the CDI algorithm outperforms the E-CDI algorithm. 
As the number of malicious UAVs increases, a shift unfolds: the E-CDI algorithm gradually overtakes the CDI algorithm, and the performance gap between them widens with the growing number of malicious UAVs.
This is because, when dealing with a small number of malicious UAVs, the CDI algorithm's scrutiny of each individual UAV is advantageous. But, as the number of malicious UAVs expands, the escalating number of malicious UAVs poses increasing feasibility challenges for the CDI algorithm. On the other hand, by prioritizing high Recall, the E-CDI algorithm aims to identify as many malicious UAVs as possible. 

\begin{figure*}[h]
\centering \epsfig{file=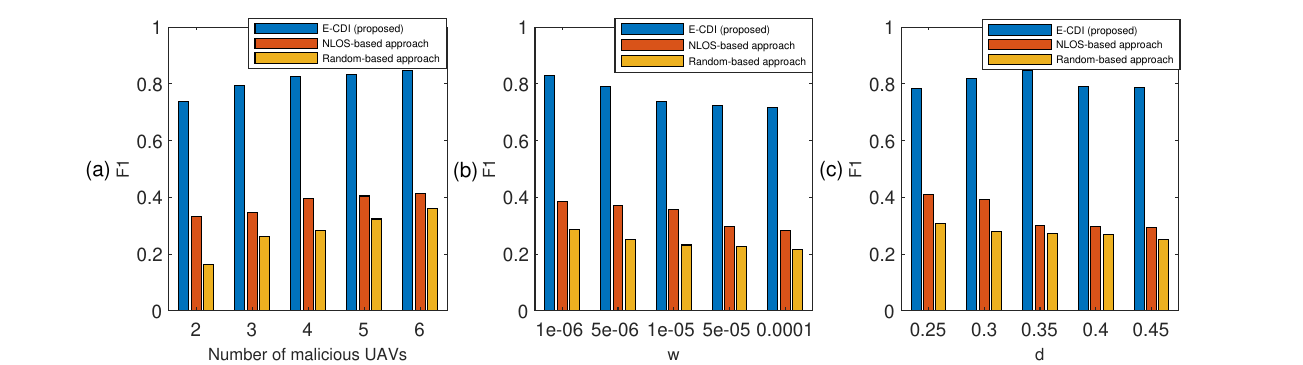, width=1.05\textwidth} 
\caption{ The performance on F1 of the proposed and baseline approaches in collusion spoofing attack scenario. (a) Under different scales of malicious UAVs. (b) Under different levels of distance measurement noise. (c) Under different levels of measurement distance.}
\label{F:combined_clusive} 
\end{figure*}

\begin{figure*}[h]
\centering \epsfig{file=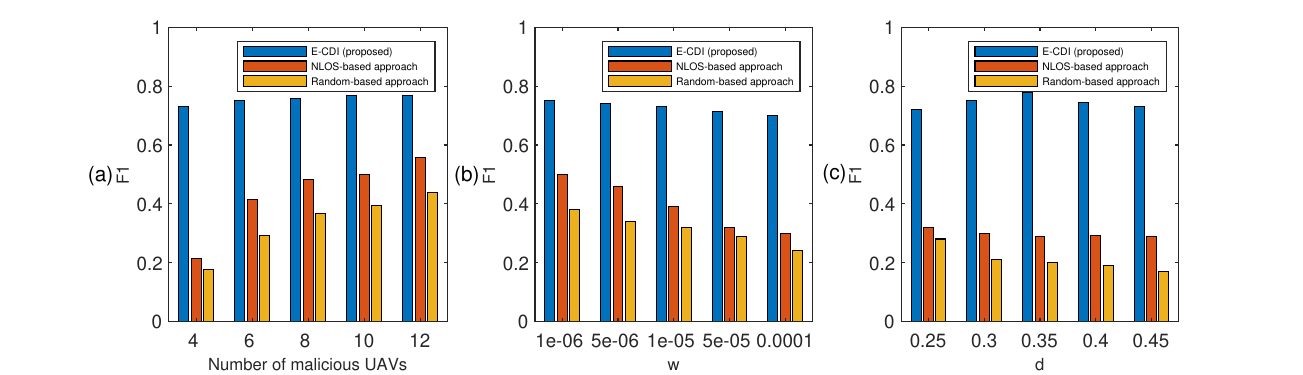, width=1.05\textwidth} 
\caption{ The performance evaluation on F1 of the proposed and baseline approaches in mixed spoofing attack scenario. (a) Under different scales of malicious UAVs. (b) Under different levels of distance measurement noise. (c) Under different levels of measurement distance.}
\label{F:combined_mixed} 
\end{figure*}

\subsection{Resistance to Collusion Spoofing Attacks 
}

In line with Fig. \ref{F:UAV2}, we delve into the evaluation of the F1 metric for the proposed E-CDI algorithm in comparison to the benchmarks. 
The CDI algorithm is not assessed, as explained in Section \ref{approach}. To maintain consistency, we generate 100 UAV swarms randomly and independently with consistent parameters with those in Fig. \ref{F:UAV2}, except for the number of malicious UAVs. Each data point represents the average of the 100 swarms.


Fig. \ref{F:combined_clusive}(a) compares the F1 score between the E-CDI algorithm and the benchmarks under a collusion spoofing attack. Unlike the declining trend observed in Fig. \ref{F:UAV_combine}(c), the proposed E-CDI algorithm exhibits an upward trend as malicious UAVs increase.
On the one hand, collusion spoofing attacks have a distinct structure compared to distributed spoofing attacks in the sense that attackers are more likely to be densely concentrated around a targeted UAV. On the other hand, the E-CDI algorithm assesses potentially malicious UAVs individually if their neighborhood is detected to be infected by malicious UAVs, which is particularly effective in the densely populated neighborhood of a collusion attack. Consequently, the E-CDI can exploit the specific characteristics of collusion spoofing attacks.

The ascending trend is also observed under the benchmark algorithms, resulting from the presence of densely concentrated UAVs, which tend to induce fewer disparities between the distance computed based on reported positions and the reported distance measurements. This, in turn, results in smaller candidate sets of malicious UAVs, which are more likely to be detected by the benchmarks.


We proceed to assess the impact of two critical parameters: distance measurement error and the range for distance measurement on the performance of the proposed E-CDI algorithm.
With the increasing distance measurement noises, a descending trend of the F1 score can be observed in Fig.~\ref{F:combined_clusive}(b), as done in Fig.~\ref{F:UAV_combine}(e). On the other hand, a descending trend is also noticed with the increasing permissible distance in Fig.~\ref{F:combined_clusive}(c), which is consistent with the observation made in Fig.~\ref{F:UAV_combine}(f).
Given that these two parameters exhibit a similar influence in both distributed and collusion scenarios, we can refer to the discussions about distributed spoofing in Section~\ref{distributed} for the sake of brevity.

\subsection{
Resistance to Mixed Spoofing Attack} 

As considered in Fig. \ref{F:UAV3}, we proceed to assess the F1 metric of the proposed E-CDI algorithm, comparing it to the benchmarks. 
A total of 100 UAV swarms are generated randomly and independently with consistent parameters with those considered in Fig. \ref{F:UAV3}, except for the number of malicious UAVs. Each data point represents the average of the 100 swarms.


In Fig. \ref{F:combined_mixed}(a), we analyze the F1 performance of the proposed E-CDI algorithm compared to the benchmarks in the context of a mixed spoofing attack. In order to conduct a fair evaluation across various numbers of malicious UAVs, we evenly distribute the malicious UAVs into two distinct groups. For instance, when dealing with six malicious UAVs, we assign three of them to execute distributed attacks. The remaining three are directed toward launching collusion attacks against a benign UAV.
It is noted that the mixed spoofing attack cannot be simply regarded as the superposition of distributed and collusion attacks. The amalgamation of these different attack types within the mixed scenario introduces substantial variations in the network structure.


Comparing the trends seen in Fig. \ref{F:combined_mixed}(a) to those in Figs.~\ref{F:UAV_combine}(c) and \ref{F:combined_clusive}(a), it is evident that the performance of the proposed E-CDI algorithm follows a similar ascending trend as observed in Fig. \ref{F:combined_clusive}(a), which is different from the trend in Fig. \ref{F:UAV_combine}(c).
As malicious UAVs increase, the network structure undergoes a notable transformation. Initially, both distributed attacks and collusion attacks contribute evenly. However, with a greater number of attackers, more malicious UAVs initially involved in distributed attacks inadvertently become participants in collusion attacks. This shift results in the gradual dominance of collusion attacks. Consequently, the performance trend observed in the mixed spoofing attack aligns with the pattern shown in Fig. \ref{F:combined_clusive}(a), although there can be a slight performance degradation for the same number of malicious UAVs.

In the case of the benchmarks, the ascending trend of their F1 scores can be attributed to the presence of densely concentrated UAVs, which tends to reduce the disparities between the distance computed based on reported positions and the reported distance measurements, as discussed earlier.


Last but not least, we assess the impact of distance measurement error and the permissible distance for distance measurement on the performance of the proposed E-CDI algorithm. 
It is noticed that Fig. \ref{F:combined_mixed}(b) yields a declining trend like the one observed in Fig. \ref{F:UAV_combine}(e), while Fig. \ref{F:combined_mixed}(c) displays a decreasing pattern like the one shown in Fig. \ref{F:UAV_combine}(f). Given that these two parameters have consistent effects under the distributed and collusion attacks, the reason underlying the observations in Figs. \ref{F:combined_mixed}(b) and \ref{F:combined_mixed}(c) can be established, as discussed in Section~\ref{distributed}.
It is worth highlighting that there exists a marginal performance decline for the same number of malicious UAVs when compared to the distributed attacks. This is due to the more intricate degradation of the network structure exacerbated by the interplay of mixed attacks.

\section{Conclusion}

In this paper, we judiciously formulated a complicated malicious UAV detection problem based on the reported positions and pairwise distance measurements of the UAVs as a localization feasibility problem.  
Then, we relied on an SDR approach to recast the formulated non-convex problem into a convex SDP, which is the key to assessing the feasibility of the reported positions and distance measurements. Moreover, we proposed two tailored iterative algorithms based on the proposed SDP approach to detect and identify malicious UAVs in UAV swarms. Extensive simulations demonstrated that the proposed algorithms can achieve superior performance to the existing benchmarks and exhibit robustness across various UAV swarms.

\bibliographystyle{IEEEtran}
\bibliography{iov2023}

\end{document}